\documentclass[preprint,12pt]{elsarticle}

\usepackage{graphicx}
\usepackage{amssymb}

\usepackage{lineno}

\begin{document}

\begin{frontmatter}


\title{Next Generation Light Sources and Future Applications}

\author{S. Kumar}

\address{Raja Ramanna Centre for Advanced Technology, India}

\begin{abstract}
The 4th generation light source has achieved tremendous success and leveraged scientific research in material science, biology and chemistry fundamentally. This paper discussed progress in LCLS as introduction and the possibility of accelerating driver beam with plasma wakefield in following section. Several potential applications of next generation light source such as imaging complex chemical reactions are listed at the end.
\end{abstract}


\end{frontmatter}


\section{Introduction}
Light source has been one of the most important tools to unveil the mystery of nature in different disciplines such as biology, material science and chemistry. The application of light is not just limited to visible light on wavelength spectrum. In 1895, R\"{o}ntgen discovered x-ray. Since then x-ray has been critical to study the atomic structure of crystals. As one of the most famous example, the double helix structure of DNA was identified from x-ray diffraction pattern. The invention of laser, ranging from infrared to UV and x-ray, especially has contributed to many scientific research. Scientists have been working on improving the brightness of light source. Synchrotron radiation which is highly collimated emission from electrons orbiting in storage ring was developed starting from 1940s based on special relativity theory and Maxwell's electromagnetic theory. 

The brilliance of light beam is a metric defined as a function of frequency given by the number of photons emitted by the source in unit time in a unit solid angle, per unit surface of the source and in a unit bandwidth of frequencies around the given one. The history of light source is a timeline of scientists improving brilliance of light source. In 1960s, storage rings were started to be employed in scientific experiments. Originally, they were designed and constructed for nuclear physics studies. The usage as light source for solid state physics were partially and parasitically. These early-days storage rings were referred as "the first generation of light source". As the potential of synchrotron was recognized, people started to construct storage rings dedicated to serve as light source which was later named "the second generation of light source". The third generation of light sources are characterized by two aspects. The first one is the progress of design and manufacturing of magnets lattice to reduce the emittance of circulating beam and enhance the collimation. The second one is the extensive use of wigglers or undulators. The brilliance of light generated in that generation were orders of magnitude higher than early generations. Examples are the Advanced Light Source in Berkeley, BESSY II in Berlin, the Advanced Photon Source in Argonne and Spring 8 in Japan.

An alternative of fourth generation light source is lower emittance storage rings. The new storage ring with low momentum compaction factor is able to produce pulse with duration of 1 ps\cite{Robin1993}. The challenges of producing higher brightness photon beam at shorter wavelength had been discussed\cite{Cornacchia1992, Laclare1996} such as the position stability and reproducibility of electron beam and single and multi-bunch instabilities, etc. Solutions were also proposed such as feedback system and high-harmonic cavities. The largest challenges among all these is so called Touschek effect\cite{Piwinski1999}. When the electron beam bunch is too short, the collisions of electrons in bunch are non-trivial which reduces beam lifetime significantly especially when beam energy is relatively low. This effect can be mitigated by increasing the vertical emittance of electron beam.

One disadvantage of storage ring light source is that electrons gain a little bit perturbation from quantum process and recoil effects each cycle. After millions of millions cycles, the aggregated perturbation is non-trivial and prevents reducing beam emittance via magnet lattice. People have also explored another path FEL since 1970s. The first FEL operated at infrared wavelengths was demonstrated in experiment\cite{Deacon1977}. Soon after that, people recognized self-amplified spontaneous emission (SASE) mode\cite{Kondratenko1980,Bonifacio1984,Murphy1985,Kim1986}.

Starting from 1990s, researchers proposed to build free electron laser operating in the far ultraviolet (UV) region which was named FLASH\cite{Ackermann2007}. The facility was commissioned in 1999. Since then, many remarkable scientific experiments had been carried out which later laid foundation for design and construction of x-ray FEL. The fundamental wavelength of FEL $\lambda_r$ is Equation \ref{eq:felwl} where $\lambda_u$ is the undulator period, $\gamma$ is electron beam's Lorentz factor, $c$ is speed of light, $\omega_r$ is the angular frequency of FEL radiation and $K$ can be written as $eB_0\lambda_u/(2\pi mc)$ is the undulator strength parameter in which $e$ is the electron unit charge, $m$ is electron static mass and $B_0$ is the peak magnetic field. 

\begin{equation}
\label{eq:felwl}
\lambda_r = \frac{\lambda_u}{2\gamma^2}(1 + \frac{K^2}{2}) = \frac{2\pi c}{\omega_r}
\end{equation}

\begin{figure}[h]
\centering\includegraphics[width=0.8\linewidth]{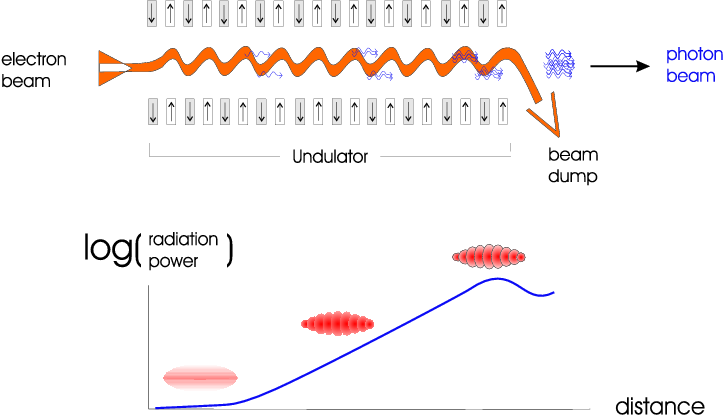}
\caption{Schematic drawing of SASE FEL. Electron beam is injected from left end of undulator\cite{Ringwald2003}, electron beam undergoes microbunching along undulator z-axis until radiation reaches saturation.}
\end{figure}

The generation of short x-ray pulse was demonstrated at SPPS project in SLAC\cite{Bentson2003}. The developments of RF and undulator technologies made it possible that using high energy and low emittance electron beam to produce much shorter wavelength FEL in a single pass through long undulator. Laboratories used photocathode rf guns and bunch length compressors to generate high peak current electron beams to produce single-pass FELs from VUV to the Angstrom range. In 1992, Claudio Pellegrini proposed to construct SASE FEL operating in 1-40 \AA range at SLAC\cite{Pellegrini1992}. In 2009, the Linac Coherent Light Source (LCLS) started to deliver first light as the world's first x-ray free electron laser. The generation of x-ray FEL requires bright electron beam and high-quality undulator. The gain of FEL amplifier in 1D is scaled by Pierce parameter.

\begin{equation}
\label{eq:pierce}
\rho = [\frac{1}{64\pi^2}\frac{I_p}{I_A}\frac{K^2(J_0(\xi) - J_1(\xi))^2\lambda_u^2}{\gamma^3\sigma_x^2}]^{1/3}
\end{equation}

where $I_p$ is the electron beam peak current, $I_A \approx 17 kA$ is the alfven current, $\sigma_x$ is the electron beam transverse size, $\xi=K^2/(4+2K^2)$. Pierce parameter indicates FEL prefers high peak current and small transverse size. The challenge of LCLS electron injector is to compress 1 nC bunch charge into a few picosecond length while limiting transverse normalized emittance to less than 1 $\mu m$. \cite{Batchelor1992,Palmer1997,Akre2008} discussed the prototype of rf gun and extensive design and engineering. The quality of x-ray FEL also put restriction on undulator. Quadrupole magnets are inserted to keep the beam size small and stable. The magnetic field strength of each undulator section has to be uniform over the entire beam line. The relative variation of $K$ should be much less than the FEL parameter $\rho$. Another challenge of undulator is the alignment. the mis-alignment tolerance should be at the level of 5 $\mu m$ over FEL gain length\cite{Emma1999}. 

\begin{table}[h]
\centering
\begin{tabular}{l l l}
\hline
\textbf{Parameters} & \textbf{Hard x-rays} & \textbf{Soft x-rays}\\
\hline
Fundamental wavelength $\lambda_r$ & 6.2-0.97 \AA & 44.3-6.2 \AA \\
Photon Energy & 2k-12.8k eV & 280-2k eV \\
Beam Energy & 6.7-16.9 GeV & 2.5-6.7 GeV \\
FEL gain length & 1.5-5.4 m & 1.0-2.6 m \\
Electron bunch charge Q & 0.15-0.25 nC & 0.15-0.25 nC \\
Injector bunch length $\sigma_{z0}$ & 650 $\mu m$ & 650 $\mu m$ \\
Undulator bunch length $\sigma_{zf}$ & 4.3-22 $\mu m$ & 6.1-35 $\mu m$ \\
Peak current $I_{pk}$ & 1.0-3.0 kA & 0.63-2.1 kA \\
Proj. emittance (injector) $\epsilon_{inj}$ & 0.45 $\mu m$ & 0.45 $\mu m$\\
Proj. emittance (undulator) $\epsilon_{und}$ & 0.5-1.6 $\mu m$ & 0.5-1.6 $\mu m$\\
\hline
\end{tabular}
\caption{Typical measured LCLS parameters with hard and soft x-rays}
\end{table}

\section{Next generation light source}
X-rays from synchrotrons and X-FELs have opened up new frontiers in many fields such as the study of protein structures, study of material defects and replication of matters near the center of the sun. However, both synchrotrons and X-FELs require the driver electron beam at least a few giga-electronvolts (GeV). The size of accelerator based on traditional RF technology has to be at least hundreds of meters to achieve that energy level due to the breakdown of RF structures' walls under larger electrical field which sets the limit of accelerating gradient to hundreds of megavolts per meter. This disadvantage constrained the growth of light source community.

In last decades, advanced acceleration technologies have been pushed forward significantly. One of the most promising candidate is laser wakefield accelerator (LWFA). A strong laser pulse injected into plasma expels plasma electrons to side while leaving ions static due to the huge mass difference between ions and electrons. The charge separation excites strong wakefields. The plasma-based accelerator's accelerating gradient is usually three orders of magnitude higher than traditional RF accelerator which has the potential to reduce the accelerator size from kilometers to table-top accelerating same energy level electron beam. The original idea of plasma-based acceleration was proposed by Tajima and Dawson in 1970s. The accelerating gradient can be estimated by plasma wave amplitude as in Equation \ref{eq:1} where $E_z$ is longitudinal component of electric field amplitude, $m_e$ is the electron static mass, $c$ is speed of light, $\epsilon_0$ is vacuum permittivity. 

\begin{equation}\label{eq:1}
E_z = \frac{m_{e}c^2}{\epsilon_0}n_{e}^{1/2} \approx 100\sqrt{n_e(cm^{-3})} 
\end{equation}

Consider the plasma density is $10^{14} cm^{-3}$, the accelerating gradient is 1 GV/m.

Experimentally, scientists have dedicated growing effort to the demonstration of laser plasma acceleration since 1990s\cite{Modena1995,Ting2005,Leemans2002}. The realization of LWFA benefited from the development of short-pulse lasers based on chirped-pulse amplification\cite{Perry1994}. Lasers in plasma experiments are mostly Ti:Sa laser with a few joules of energy, sub-hundred femtoseconds of pulse duration and 800 nm of centered wavelength. Nonetheless, the quality of accelerated electron beam is not satisfactory with most of electrons distributed at the low energy end until three independent groups published their impressive results of accelerated beam with $\sim$100 pC at energy of $\sim$100 MeV\cite{Faure2004,Geddes2004,Mangles2004}.

\begin{figure}[h]\label{fig:lwfa_xfel}
\centering\includegraphics[width=0.8\linewidth]{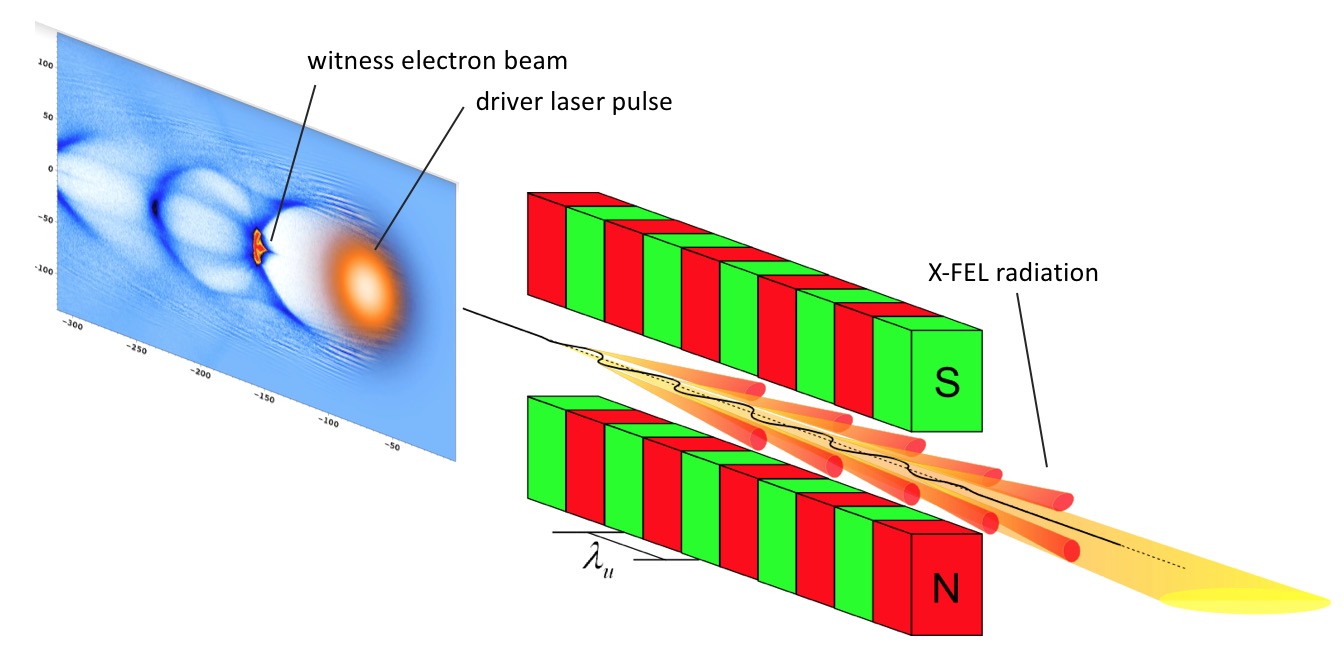}
\caption{Schematic of X-FEL light source with LWFA-based injector. Laser wakefield accelerator delivers electron beam of low emittance and high peak current. The electron beam from LWFA is later injected into undulator to generate X-FEL.}
\end{figure}

To support the rapid growth of light source demands, scientists have put forward the design of table-top laser-based X-FEL with the demonstration of quasi-monoenergetic GeV electron beam generated from LWFA as shown in Figure \ref{fig:lwfa_xfel}. Plasma environment can be created by gas jet or capillary. Compared to gas jet\cite{Couperus2016,Krishnan2009}, laser can be guided beyond Rayleigh length and thus witness beam has longer accelerating distance. The production of high-quality electron beam with 1 GeV energy by a 40 TW peak-power, 38 fs duration laser pulse in a 3.3-cm-long gas-filled capillary discharge waveguide has been demonstrated in Lawrence Berkeley National Laboratory\cite{Leemans2006}. The 30 pC witness beam charge has not reached the expectation of 1 nC. Therefore, scientists have proposed employment of higher gas density and Petawatt ultrashort ($\sim$5 fs) laser\cite{Gruner2007}. The feature of this scheme is that as laser transverse size grows, the scenario transits from bubble regime to blow-out regime adiabatically.

In general, LWFA nowadays can deliver electron beams from hundreds of MeV to 1 GeV with relative energy spread of the order of 1\% and normalized emittance of 
1 mm.mrad thus peak current of a few kA\cite{Malka2008,Rechatin2009,Fritzler2004,Brunetti2010}. As a comparison, the typical electron beam of current operating X-FEL facility has 0.1-1 nC charge, 1 mm.mrad emittanceand 0.01\% energy spread at a few GeV. LWFA has undesired larger energy spread and divergence which cannot trigger direct FEL amplification. Adequate efforts have been dedicated to manipulate beam to make it qualified for FEL amplification. For example, installing a quadrupole close to the source can mitigate the growth of emittance and bunch duration\cite{Anania2010}. To address the energy spread issue, one can implement a chicane which is composed of four dipole magnets\cite{Maier2012}.

A novel scheme "Trojan horse" plasma wakefield was proposed in 2012\cite{Hidding2012,Hidding2013}. In this scenario, a electron driver bunch drives a wake in gas mixture by ionizing the low-ionizing-threshold gas such as Li or $\mathrm{H_2}$ and creates a blow-out regime. Following the driver beam, a synchronized ultrashort ($\sim$50 fs) laser pulse (within $\sim 100 fs$) is injected and focused to ionize high-ionizing-threshold gas such as He to release witness beam electrons. These electrons will be trapped on the condition that the ionization laser is injected into appropriate phase. As the electrons released latter will be beam front, the scenario has self-compression feature in longitudinal direction. Transversally, initial emittance due to laser tunnel ionization is trivial. According to simulation in \cite{Xi2013}, the normalized emittance of witness from this scheme can be 0.01 mm.mrad and peak current can be over 100 kA. Proof-of-concept experiment has been designed and carried out at FACET in SLAC\cite{Wittig2015,Wittig2016,Manahan2015,Manahan2016,Bernhard2014,Hidding2014,Knetsch2014}. This scheme has the potential to deliver electron beam qualified for X-FEL directly and produce even shorter wavelength FEL.

\begin{figure}[h]
\label{fig:betatron}
\centering\includegraphics[width=0.7\linewidth]{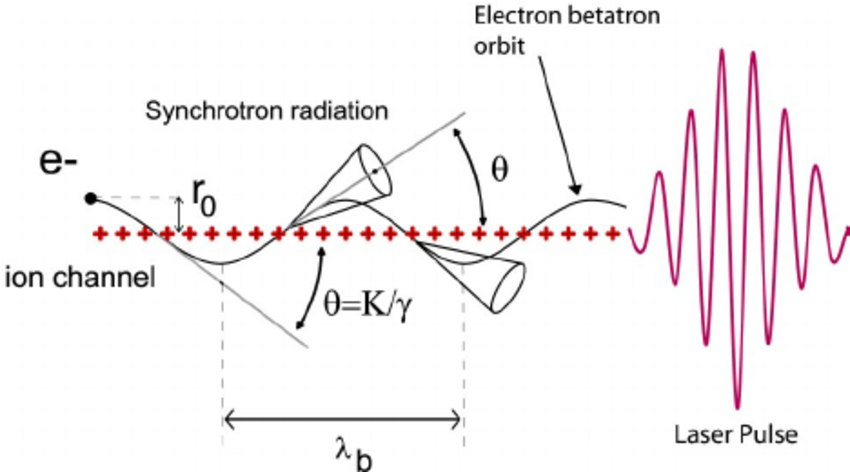}
\caption{An off-axis injected single electron emits radiation in ion channel created by laser pulse. Betatron wavelength is denoted as $\lambda_b$. $\theta$ is angle corresponding to the peak angular deflection of the electron which represents the opening angle where betatron radiation is confined.}
\end{figure}

Another method to generate radiation based on LWFA is betatron radiation as shown in Figure \ref{fig:betatron}. If the witness beam is injected slightly off the axis intentionally, it will experience transverse restoring force towards the axis in ion column. Thus electrons has oscillating motion in transverse while accelerated in longitudinal. radiation photons are emitted along the sinusoidal trajectory. Betatron radiation is very similar to synchrotron X-rays except in synchrotron  electrons' trajectory is bent by periodic magnetic structure while in betatron radiation source, periodic magnetic structure is replaced by plasma wave. Both sources emit continuous X-ray spectral range and in a narrow forwarded cone.

\section{Applications}
Next generation light source will provide brighter and higher repetition rate X-ray laser which has unique capabilities to lead to a new understanding in many reactions by enabling cinematic imaging of dynamics of systems.

The detail processes how to convert water and carbon dioxide into organic compounds and oxygen are not absolutely clear yet. Next generation light source has an opportunity to unveil the mystery of natural photosynthetic reactions. The ultrafast laser pulse can track chemical process from transferring of electrons between molecules in quantum mechanics to slow reactions related to solar energy. This study also provides insightful guidance to improve conversion rate of solar energy to clean energy\cite{Lewis2006,Ferreira2004}.

New light source will also be instrumental to combustion science which is a complex mixing reaction of chemical and turbulent fluid mechanics\cite{Dec1997}. X-ray from light source can image chemistry and physics governing the whole process from fuel sprays to gas combustion to particulate formation and evolution. In particular, the high power and coherence and narrow bandwidth X-ray can stimulate Raman spectroscopies with orders of magnitude higher efficiency.

Another application of next generation light source is to visualize the catalytic conversion processes so that researchers have a better understanding of reaction pathways and catalytic conversion efficiency. In detail, these electro- and photochemical processes can be analyzed by pump-probe X-ray absorption spectroscopy and time-resolved photoemission spectroscopy. This could advance the search for optimal catalyst to split $\mathrm{H_2O}$ molecules to generate clean Hydrogen energy significantly.

\section{Summary}
In summary, light sources have contributed remarkably to fields like medical, biology and material science in aspects such as imaging, structural determination and spectroscopy. As research in advanced accelerators especially in plasma-based wakefield accelerators has achieved tremendous progress in last decades. This could bring another big leap towards higher brightness, even shorter wavelength and higher repetition rate X-rays. The next generation light source will provide unprecedented capabilities to study ultrashort time-scale reactions and spark revolutionary progress in industries.

\bibliographystyle{unsrt}
\bibliography{reference}

\begin{thebibliography}{10}

\bibitem{Robin1993}
D.~Robin et~al.
\newblock {\em Phys. Rev. E}, 48:2149, 1993.

\bibitem{Cornacchia1992}
H.~Winick M.~Cornacchia.
\newblock {\em Workshop on Fourth Gen. Light Sources}, 1992.

\bibitem{Laclare1996}
J.-L.~Laclare et~al.
\newblock {\em Workshop on Fourth Generation Light Sources}, 1996.

\bibitem{Piwinski1999}
A.~Piwinski et~al.
\newblock {\em arxiv}, 9903034, 1999.

\bibitem{Deacon1977}
D.~Deacon et~al.
\newblock {\em Phys. Rev. Lett.}, 38:892, 1977.

\bibitem{Kondratenko1980}
A.~Kondratenko et~al.
\newblock {\em Part. Accel.}, 10:207, 1980.

\bibitem{Bonifacio1984}
R.~Bonifacio et~al.
\newblock {\em Opt. Commun.}, 50:373, 1984.

\bibitem{Murphy1985}
J.~Murphy et~al.
\newblock {\em J. Opt. Soc. Am. B}, 2:259, 1985.

\bibitem{Kim1986}
K.-J.~Kim et~al.
\newblock {\em Phys. Rev. Lett}, 57:1871, 1986.

\bibitem{Ackermann2007}
W.~Ackermann et~al.
\newblock {\em Nat. Photonics}, 1:336, 2007.

\bibitem{Ringwald2003}
A.~Ringwald et~al.
\newblock {\em arxiv:hep-ph}, 0304139, 2003.

\bibitem{Bentson2003}
L.~Bentson et~al.
\newblock {\em Nuclear Instruments and Methods in Physics Research A},
  507:205--209, 2003.

\bibitem{Pellegrini1992}
C.~Pellegrini et~al.
\newblock {\em 4th Generation Light Source Workshop proceedings}, 92-02:345,
  1992.

\bibitem{Batchelor1992}
K.~Batchelor et~al.
\newblock {\em Nucl. Instrum. Methods Phys. Res., Sect. A}, 318:372, 1992.

\bibitem{Palmer1997}
D.~Palmer et~al.
\newblock {\em Proceedings of the 1997 Particle Accelerator Conference},
  3:2687--2689, 1997.

\bibitem{Akre2008}
R.~Akre et~al.
\newblock {\em Phys. Rev. ST Accel. Beams}, 11:030703, 2008.

\bibitem{Emma1999}
P.~Emma et~al.
\newblock {\em Nucl. Instrum. Methods Phys. Res., Sect. A}, 429:407, 1999.

\bibitem{Modena1995}
A.~Modena et~al.
\newblock {\em Nature}, 377:606, 1995.

\bibitem{Ting2005}
A.~Ting et~al.
\newblock {\em Phys. Plasmas}, 12:010701, 2005.

\bibitem{Leemans2002}
W.~P.~Leemans et~al.
\newblock {\em Phys. Rev. Lett.}, 89:174802, 2002.

\bibitem{Perry1994}
M.~D.~Perry et~al.
\newblock {\em Science}, 264:917, 1994.

\bibitem{Faure2004}
J.~Faure et~al.
\newblock {\em Nature}, 431:541, 2004.

\bibitem{Geddes2004}
G.~G. R.~Geddes et~al.
\newblock {\em Nature}, 431:538, 2004.

\bibitem{Mangles2004}
S.~P. D.~Mangles et~al.
\newblock {\em Nature}, 431:535, 2004.

\bibitem{Couperus2016}
J.~P.~Couperus et~al.
\newblock {\em Nucl. Instrum. Methods Phys. Res., Sect. A}, 830:504--509, 2016.

\bibitem{Krishnan2009}
M.~Krishnan et~al.
\newblock {\em AIP Conference Proceedings}, 1086:264, 2009.

\bibitem{Leemans2006}
W.~P.~Leemans et~al.
\newblock {\em Nature Physics}, 2:696--699, 2006.

\bibitem{Gruner2007}
F.~Gruner et~al.
\newblock {\em Appl. Phys. B}, 86:431--435, 2007.

\bibitem{Malka2008}
V.~Malka et~al.
\newblock {\em Nat. Phys.}, 4:447--53, 2008.

\bibitem{Rechatin2009}
C.~Rechatin et~al.
\newblock {\em Phys. Rev. Lett.}, 102:164801, 2009.

\bibitem{Fritzler2004}
S.~Fritzler et~al.
\newblock {\em Phys. Rev. Lett.}, 92:165006, 2004.

\bibitem{Brunetti2010}
E.~Brunetti et~al.
\newblock {\em Phys. Rev. Lett.}, 105:215007, 2010.

\bibitem{Anania2010}
M.~P.~Anania et~al.
\newblock {\em SPIE Conference Proceedings}, 7359:735916, 2009.

\bibitem{Maier2012}
A.~R.~Maier et~al.
\newblock {\em Phys. Rev. X}, 2:031019, 2012.

\bibitem{Hidding2012}
B.~Hidding et~al.
\newblock {\em Phys. Rev. Lett.}, 108:035001, 2012.

\bibitem{Hidding2013}
B.~Hidding et~al.
\newblock {\em AIP Conference Proceedings}, 1507:570, 2013.

\bibitem{Xi2013}
Y.~Xi et~al.
\newblock {\em Phys. Rev. ST Accel. Beams}, 16:031303, 2013.

\bibitem{Wittig2015}
G.~Wittig et~al.
\newblock {\em Phys. Rev. ST Accel. Beams}, 18:081304, 2015.

\bibitem{Wittig2016}
G.~Wittig et~al.
\newblock {\em "Nuclear Instruments and Methods in Physics Research Section A:
  Accelerators, Spectrometers, Detectors and Associated Equipment"}, 829:83 --
  87, 2016.

\bibitem{Manahan2015}
G.~G. Manahan and O.~Karger et~al.
\newblock Towards first realization of trojan horse underdense photocathode
  plasma wakefield acceleration.
\newblock {\em 42nd EPS Conference on Plasma Physics}, page P2.222, 2015.

\bibitem{Manahan2016}
G.~G. Manahan and A.~Deng et~al.
\newblock {\em Phys. Rev. Accel. Beams}, 19:011303, 2016.

\bibitem{Bernhard2014}
B.~Hidding et~al.
\newblock Tunable electron multibunch production in plasma wakefield
  accelerators.
\newblock {\em arXiv}, 1403.1109, 2014.

\bibitem{Hidding2014}
B.~Hidding et~al.
\newblock {\em Journal of Physics B: Atomic, Molecular and Optical Physics},
  47:234010, 2014.

\bibitem{Knetsch2014}
A.~Knetsch et~al.
\newblock Downramp-assisted underdense photocathode electron bunch generation
  in plasma wakefield accelerators.
\newblock {\em arXiv preprint}, 1412.4844, 2014.

\bibitem{Lewis2006}
N.~S.~Lewis et~al.
\newblock {\em Proceedings of the National Academy of Sciences of the United
  States of America}, 103, 2006.

\bibitem{Ferreira2004}
K.~N.~Ferreira et~al.
\newblock {\em Science}, 303, 2004.

\bibitem{Dec1997}
J.~E.~Dec et~al.
\newblock {\em SAE Technical Paper}, page 970873, 1997.

\end{thebibliography}

\end{document}